\documentclass[12pt,a4paper]{conference}

\usepackage{fancyhdr}
\usepackage{graphicx,amsmath,amssymb,cite}
\usepackage{multind}
{\vskip 1.0em
\framebox[\columnwidth][r]{%
\begin{minipage}[c]{\columnwidth}%
\vspace{-1.0em}%
#1%
\end{minipage}}}
{\vskip 1.0em}

\newcommand{\gonn}{\ensuremath{g_{\omega N\!N}}}
\newcommand{\grnn}{\ensuremath{g_{\rho N\!N}}}
\newcommand{\gopr}{\ensuremath{g_{\omega\pi\rho}}}

\newcommand{\grpp}{\ensuremath{g_{\rho\pi\pi}}}

\newcommand{\Gbar}{\overline{\Gamma}}

\newcommand{\gnon}{\ensuremath{\gamma N\to \omega N}}
\newcommand{\gpop}{\ensuremath{\gamma p\to \omega p}}
\newcommand{\pnon}{\ensuremath{\pi N\to \omega N}}

\newcommand{\pmpon}{\ensuremath{\pi^- p\to \omega n}}

\newcommand{\krho}{\ensuremath{\kappa_\rho}}
\newcommand{\komg}{\ensuremath{\kappa_\omega}}

\newcommand{\dsdo}{{\frac{d\sigma}{d\Omega}}}

\newcommand{\pn}{\ensuremath{\pi N}}
\newcommand{\en}{\ensuremath{\eta N}}
\newcommand{\pD}{\ensuremath{\pi \Delta}}
\newcommand{\sn}{\ensuremath{\sigma N}}
\newcommand{\rn}{\ensuremath{\rho N}}
\newcommand{\on}{\ensuremath{\omega N}}
\newcommand{\ppn}{\ensuremath{\pi\pi N}}

\newcommand{\beq}{\begin{equation}}
\newcommand{\eeq}[1]{\label{#1}\end{equation}}
\newcommand{\eeqn}{\end{equation}}

\newcommand{\beqa}{\begin{eqnarray}}
\newcommand{\eeqa}[1]{\label{#1}\end{eqnarray}}
\newcommand{\eeqan}{\end{eqnarray}}

\let\bar=\overbar

\newcommand{\ket}[1]{\left| {#1} \right\rangle}

\newcommand{\Dslash}{\not{\hbox{\kern-4pt $D$}}}
\newcommand{\dslash}{\not{\hbox{\kern-2pt $\del$}}}

\newcommand{\msb}{{\bar{\ssstyle M \kern -1pt S}}}

\begin{document}
%%%%%%%%%%%%%%%%%%%%%%%%%%%%%%%%%%%%%%%%%%%%%%%%%%%%%%%%%%%%%%%%%%%%%%%

\Chapter{Dynamical coupled channel approach to omega meson production}
           {Dynamical coupled channel omega production}{M.\ Paris}
\vspace{-6 cm}\includegraphics[width=6 cm]{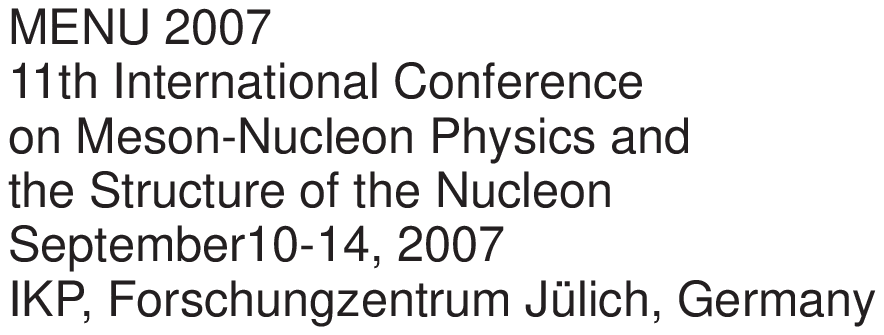}
%\bigskip\bigskip
\vspace{4 cm}

\addcontentsline{toc}{chapter}{{\it M.\ Paris}} \label{authorStart}
%%%%%%%%%%%%%%%%%%%%%%%%%%%% NEW SWITCHES %%%%%%%%%%%%%%%%%%%%%%%%%%%%%%

\begin{raggedright}

{\it M.\ Paris
\index{author}{Paris, M.}\\
Excited Baryon Analysis Center \\
Theory Group\\
Thomas Jefferson National Accelerator Facility\\
12000 Jeffereson Avenue MS 12H2\\
Newport News, Virginia 23606}
\bigskip\bigskip

%%%%%%%%%%%%%%%%%%%%%%%%%%%%%%%%%%%%%%
%%%%%%%%%%%
%%%%%%%%%%%  Repeat for second author
%%%%%%%%%%%
%%%%%%%%%%%%%%%%%%%%%%%%%%%%%%%%%%%%%%
\end{raggedright}

\begin{center}
\textbf{Abstract}
\end{center}
The dynamical \index{subject}{dynamical}
coupled channel \index{subject}{coupled channel}
approach is applied
to study the $\omega$--meson \index{subject}{omega meson}
production induced by pions and photons
scattering from the proton. \index{subject}{photoproduction}
The parameters of the model are fixed in a
two-channel (\on,\pn) calculation for the non-resonant and resonant 
contributions to the $T$ matrix by fitting the available unpolarized 
differential cross section data. The polarized photon beam asymmetry is 
predicted and compared to existing data.

\section{Introduction}
Nucleon resonances \index{subject}{nucleon resonance}
are thought to play a decisive role in reactions of
strong, electromagnetic and weak probes on nucleons.
The extent to which nucleon resonances determine both unpolarized and
polarized observables \index{subject}{polarized observable}
remains an open question in the second and third 
resonance regions from center-of-mass (COM) energies, $E$ in the 
range 1.4 GeV $\le E \le 2.0$ GeV. A model determination of the $T$
matrix consistent with the observed meson production 
\index{subject}{meson production} data in this 
kinematic regime seeks to resolve the resonance structure of 
the nucleon. The determination of the resonance spectrum will yield
insight into fundamental aspects of quantum chromodynamics, such as 
confinement\index{subject}{QCD}.

In the second and third resonance regions,\index{subject}{second resonance
region} \index{third resonance region}
channels such as \en, \pD, \rn, \sn, \on\  and \ppn\ become important.
Matsuyama, Sato, and Lee (MSL) \cite{MSL} have developed a dynamical 
coupled channel formalism to handle any number of channels with up to
three particles in intermediate and final states. In this exploratory 
study, we consider the $\omega$ meson
production reactions, \pmpon\ and \gpop, in a two-coupled channel \pn,
\on\ formalism. A full calculation incorporating the effect of stable
\en\ and unstable channels \pD, \rn, \sn\ is being pursued presently.

In this study our objective is to show that it is
possible (when the $\omega$ production is extended to include the
four additional channels \en, \pD, \rn, \sn) to predict polarized observables
from the model resulting from fits to the unpolarized differential cross
section data.

We will briefly describe the MSL model theory for the 
two-channel model for \pn\ and \on and present results of the fit.

\section{Model reaction theory\label{sec:model}}
The $T$ matrix for \gnon\ and \pnon\ 
is separated into non-resonant, $t$ and resonant,
$t^R$ terms
\begin{align}
T(E) = t(E) + t^R(E),
\end{align}
where $E$ is the scattering energy of the particles in the center-of-mass
frame. The non-resonant contribution is a smooth, regular operator function 
of the energy while the resonant term is meromorphic in the energy. No 
additonal
assumption is made about the relative size of the contributions of
these terms.

The non-resonant contribution, $t(E)$ to the full transition matrix
satifies the (relativistic) Lippmann-Schwinger (LS) equation
\index{subject}{Lippmann-Schwinger}
\begin{align}
\label{eqn:LS}
t(E) &= [1+t(E)G_0(E)]v
\end{align}
At leading order in the coupling constants of
the Lagrangian the kernel, $v$ is given by the Born amplitudes
for pion production and 
photoproduction amplitudes. We neglect the contribution of the $\eta$,
$a_0$, and $f_2$ in the non-resonant terms.

The kernel depends on coupling and cutoff parameters which are 
varied (along with the resonance parameters, see below) to fit the observed 
data. The varied parameters are
\begin{align}
\label{eqn:v-vary}
v &= v(\grnn,\krho,\grpp,\gonn,\komg,\gopr,\Lambda_{\pi NN},\Lambda_{\rho NN},
\Lambda_{\rho\pi\pi},\Lambda_{\omega NN},\Lambda_{\omega\pi\rho}).
\end{align}
The remaining parameters are fixed at the SL \cite{Sato:1996gk} values.
Form factors are assumed at each vertex. The gauge invariance is ensured 
only at the Born amplitude level. Inclusion of coupled channel and 
rescattering effects violates the gauge invariance. This is obviously an 
unsatisfactory aspect of the model which is hoped to nevertheless be 
useful in analyzing meson production reactions.

All particles, except the resonances $N^*$, are assumed to stable. In
particular, we neglect the width of the $\omega$ meson
$\Gamma_\omega=8.5(1)$ MeV.

The resonant contribution, $t^R(E)$ to the scattering matrix is given as
\begin{align}
t^R(E) &= \Gbar(E) \frac{1}{E-H_0-\Sigma(E)} \Gbar(E),
\end{align}
where $H_0\ket{N^*}=M^{(0)}_{N^*}\ket{N^*}$ and
$\Gbar(E)$ is the dressed vertex operator and $\Sigma(E)$ the
resonance self-energy.

The resonance parameters which can be tuned to fit the experimental data
are the bare masses $M_{N^*}^{(0)}$, 
the amplitudes $G^{JT}_{LSMB}$, $A^{JT}_{\lambda,T_{N,z}}$, and the
cutoffs $\Lambda^{JT}_{LS}$, $\Lambda^{JT}_{\lambda}$. In practice we
only use two cutoffs. One for all the strong vertices,
$\Lambda^{JT}_{LS}=\Lambda_M$ and one for the electromagnetic vertices
$\Lambda^{JT}_{\lambda}=\Lambda_\gamma$.

\section{Results and discussion\label{sec:results}}
\begin{figure}
\begin{center}
\includegraphics[width=200pt,keepaspectratio,angle=0,clip]{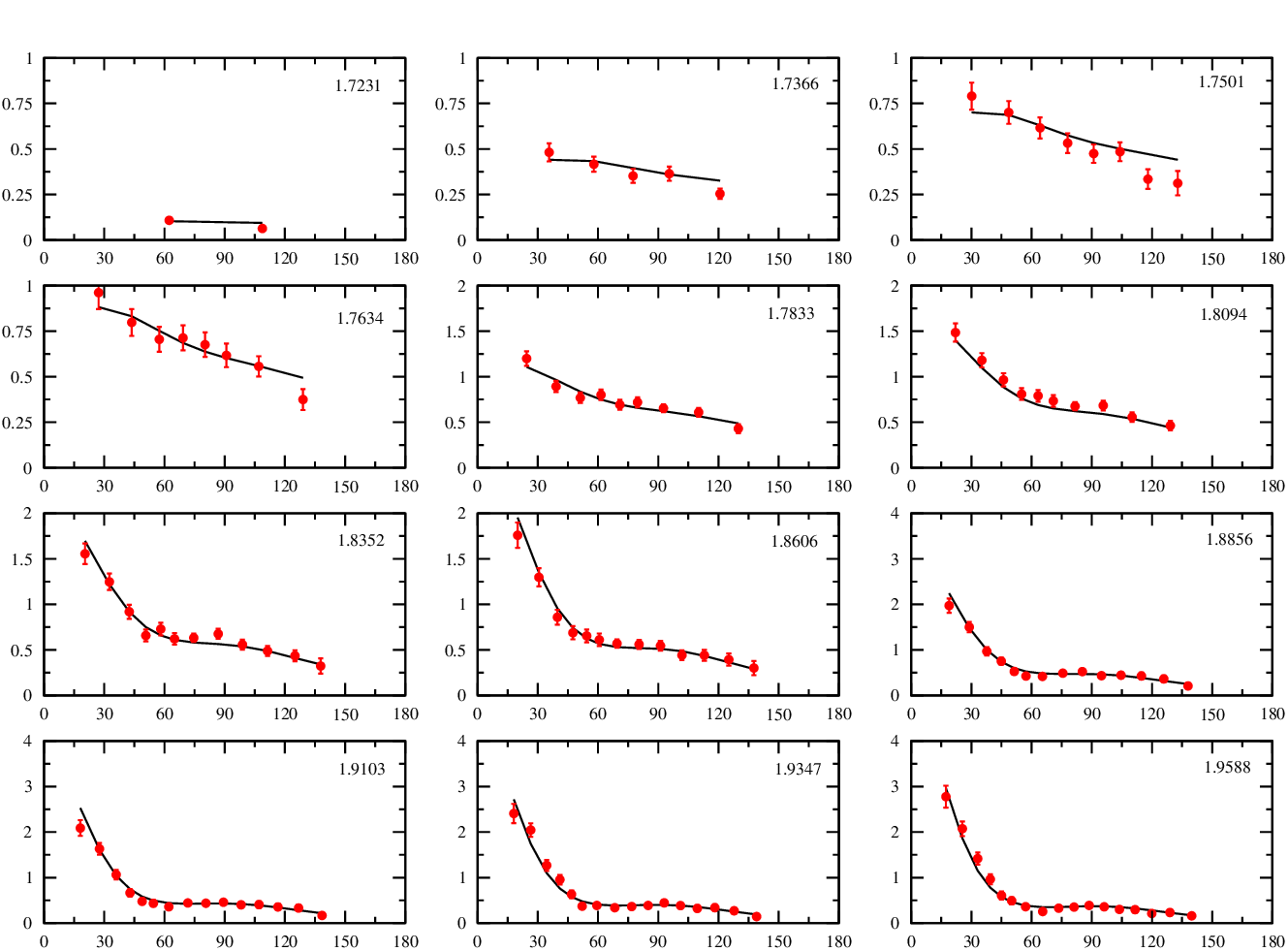}
\caption{\label{fig:gdxs}Observed unpolarized differential cross section
\cite{Barth:2003kv} (circles) for $\gpop$ compared to calculated values.
The differential cross section, $\dsdo$ in $\mu b/\mbox{sr}$ is plotted
against the $\omega$ emission angle, $\theta$ in the center-of-mass frame.
Each panel shows the cross section for given center-of-mass energy, $E$ in
GeV in the upper-right corner.}
\includegraphics[width=150pt,keepaspectratio,angle=0,clip]{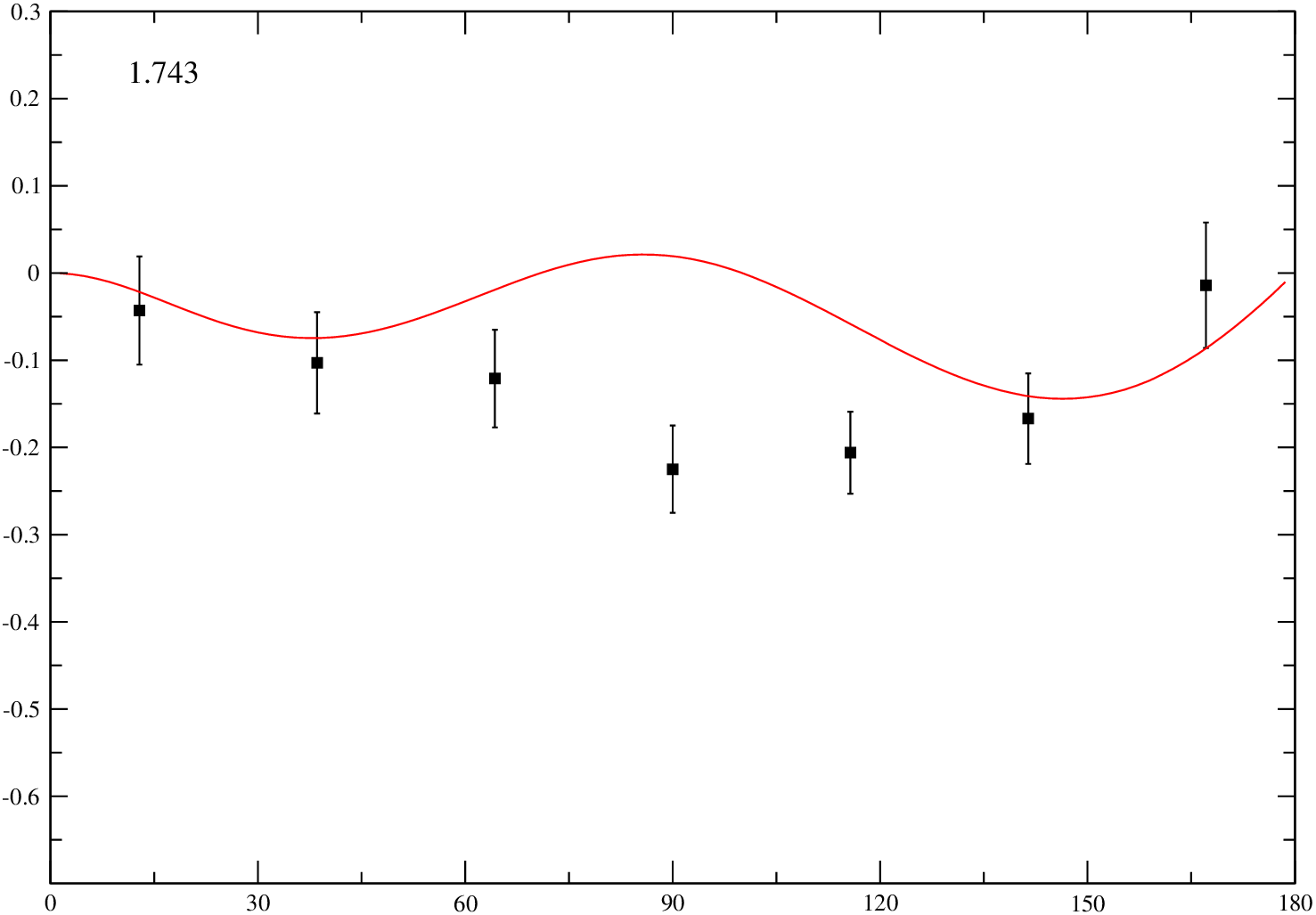}
\caption{\label{fig:sigb}Photon beam asymmetry compared to data
from GRAAL \cite{Ajaka:2006bn}
for values of COM energy shown in the lower left-hand
corner of each panel. Solid curves are for the full calculation. At energy
$E=1.743$ GeV the sensitivity to the resonance contribution is studied
(see text).}
\end{center}
\end{figure}

The fit to the \gpop\ data is shown in Fig.\eqref{fig:gdxs}. The
simultaneous fit (not shown here) to the \pmpon\ data is of similar 
quality. In particular, this fit near threshold is consistent with 
data. The behavior near threshold appears to be described well
in the current coupled channel approach \cite{Penner:2001fv}. The prediction
for the linearly polarized photon beam asymmetry $\Sigma(\theta;E)$ is
shown for $E=1.743$ GeV in Fig.\eqref{fig:sigb}. Although the predicted
$\Sigma(\theta;E)$ is not consistent with the data, the size and sign are
correct. It is hoped that the effect of other channels such as $\pD,\rn$,
etc.\ and possibly other resonances will help to improve the agreement.

%\begin{acknowledgments}
This work is supported by the U.S.\ Department of Energy, Office of
Nuclear Physics Division under contract No. DE-AC02-06CH11357, and
contract No. DE-AC05-060R23177 under which Jefferson Science Associates
operates Jefferson Lab, and by the Japan Society for the Promotion of
Science, Grant-in-Aid for Scientific Research\copyright 15540275.
%\end{acknowledgments}

%\bibliography{master}

%\end{document}

%\begin{thebibliography}{000} %for 3 digits
%\begin{thebibliography}{00}  %for 2 digits

%%%%%%%%%%%%%%%   Author and Subject Index
%\printindex{author}{Author Index}
%\blankpage

%\printindex{subject}{Subject Index}
% \blankpage

\end{document}